# ON THE DECREASE OF SUNSPOT ACTIVITY AND ABSENCE OF SEVERE SPACE WEATHER CONDITIONS DURING THE MUTIPLE PLANETARY CONJUNCTION PERIODS OF 1850-2000 AD

Nisha.N.G[1] and T.E.Girish[2*]

1 Department of Physics,Government College for Women,Trivandrum 695014, India

2 Department of Physics,UniversityCollege,Trivandrum 695034, India

*Corresponding Author, E mail : tegirish5@yahoo.co.in

We have studied solar activity and space weather conditions during multiple planetary conjunction periods ( MPC) as observed from Earth during the years 1850-2000 AD. Significant decreases in sunspot activity and solar 10.7 cm radio flux and absence of severe space weather conditions is found during most of these MPC . Possible planetary influences in longitudinal organisation of sunspot activity is suggested by these results. The success of major space missions ( lunar and human) during MPC is also noticed during 1957-2000 AD. The applications of the above results include: short term space mission planning and inference/prediction of solar-terrestrial conditions during multiple planetary conjunction periods.

## 1.Introduction

The study of planetary effects on sunspot activity is more than 150 years old (Charbonneau, 2002; Charbonneau, 2012 ) Recent research efforts focus on the understanding of possibilities of maintenance of 11 year and secular solar cycles by the planetary dynamics in the solar system . One of the earliest studies revealed that the average size of a spot would appear to attain its maximum on that side of the Sun which is turned away from Venus or from Mercury, and to have its minimum in the neighborhood of Venus or Mercury (De La Rue, et.al., 1872). Shuster (1911) studied how relative helio longitudinal organisation of sunspot activity is affected by different planets. Jose (1965) studied the importance of the periodicities involved in the motion of the Sun relative to the centre of mass of solar system and operation of the solar dynamo. Ambroz ( 1971) studied the longitudinal organization of solar activity in association with the variations in planetary tidal forces. Hung ( 2007) observed an 11 year cycle of the alignment of Venus, Earth and Jupiter similar to the 11.1 year Schwabe's sunspot cycle, from the daily planet positions during the period from 1840 – 2000 AD (Hung, 2007). Okhlopkov (2013) studied the role of conjunctions and oppositions of Venus, Earth and Jupiter in association with the sunspot cycles. Several recent studies put forward theoretical mechanisms (Ian Edmonds,2016;Stefani et al,2019) to support the idea that the tidal and angular momentum effects of planets could be amplified in the solar interior (Scafetta, 2012; Wolff and Patrone, 2010; Abreu, et al., 2012)

Planetary conjunctions are studied in all ancient civilizations like India, Greece and China. Observations of multiple planetary conjunctions were used by ancient astronomers for making calendars ( Chambers,1889;Meis and Meeus,1994 ). The conjunction of all visible planetary objects in the beginning of the zodiac during the mid night of February 18, 3102 BC is considered as the beginning of *Kali yuga* in India. The number of days elapsed from this date is considered in ancient India (similar to Julian day in modern astronomy) to record major events in history. The 60 year cycle in the re appearance of conjunctions of large planets Jupiter and Saturn in a particular zodiacal sign was given importance in this connection. It is believed that planetary conjunctions can cause natural hazards like floods and earth quakes. However there are no explicit studies in the modern context on the different solar-terrestrial aspects of multiple planetary conjunctions in spite of the vast literature existing on the planetary effects on solar activity.

In this paper, first we will specify a criteria to identify multiple planetary conjunctions (MPC) involving three planets apart from Earth using a geocentric coordinate system. The daily sunspot activity, daily sunspot area and daily solar 10.7 cm flux variations during the MPC periods are then studied for the years 1850-2000 AD depending on the availability of relevant data. The occurrence of extreme space weather events if any during MPC periods is then studied. The data on successful space missions during planetary conjunctions period is also investigated. The possible applications of the results of present studies will be discussed. This paper is a continuation of our earlier studies on this topic ( Nisha, 2003; Girish and Nisha,2012; Nisha,2015).

## 2. Criteria and Identification of Multiple Planetary Conjunctions

The criteria chosen for the identification of Multiple Planetary Conjunctions (MPC) in this study are :

( a) Conjunctions between five visible planets of our solar system, viz., Mercury, Venus, Mars, Jupiter and Saturn with the Sun (projection of position of Earth) are considered.

(b) The co-ordinate system selected to identify the position of the Sun and planets is geocentric co-ordinate system.

(c) Geocentric longitudes of the planets and Sun in conjunction should be within 20°.

(d) The minimum number of planets in conjunction should be three of which one should be either Jupiter or Saturn.

We have identified such 128 Multiple Planetary conjunctions satisfying the above criteria during the years 1850-2000 AD using the published values of the geocentric longitudes ( Swamikannu Pillai ,1985) of visible planets (Mercury,Venus,Mars,Jupiter and Saturn ) and the Sun ( projection of earth's orbital motion)

## 3 On the decrease of sunspot activity and solar 10.7cm radio flux during Multiple Planetary conjunctions during 1850-2000 AD

### 3.1 Selected Examples of sunspot activity decrease

Both international sunspot number (pre-revised) and Greenwich sunspot area ( NGDC website) are considered as excellent indicators of solar activity . The variations of daily sunspot number and daily sunspot area in and around the periods of planetary conjunctions were analyzed graphically for the entire 128 conjunctions identified. Typical examples distinct sunspot activity decreases during MPC are shown in Figure 1. In 1916 July a five planet conjunction, in 1962 February a six planet conjunction and in 2000 May a five planet conjunction occurred. In each figure the period of planetary conjunction was marked as thin vertical lines and the thick vertical arrow indicate the date of closest conjunction (CCD), the day on which the planetary longitudes have a minimum separation. The planets in conjunction and their respective geocentric longitudes are given in right column of each Figure. The good correlation between the variation of both sunspot number and sunspot area is evident for each of the above MPC period.

The variations of the daily solar 10.7cm radio flux ( NGDC website) during the 1962 Februrary MPC is shown in Fig 2. Solar radio flux also shows distinct decreases during planetary conjunction periods which is found to be correlated with sunspot number decreases as shown in this Figure.

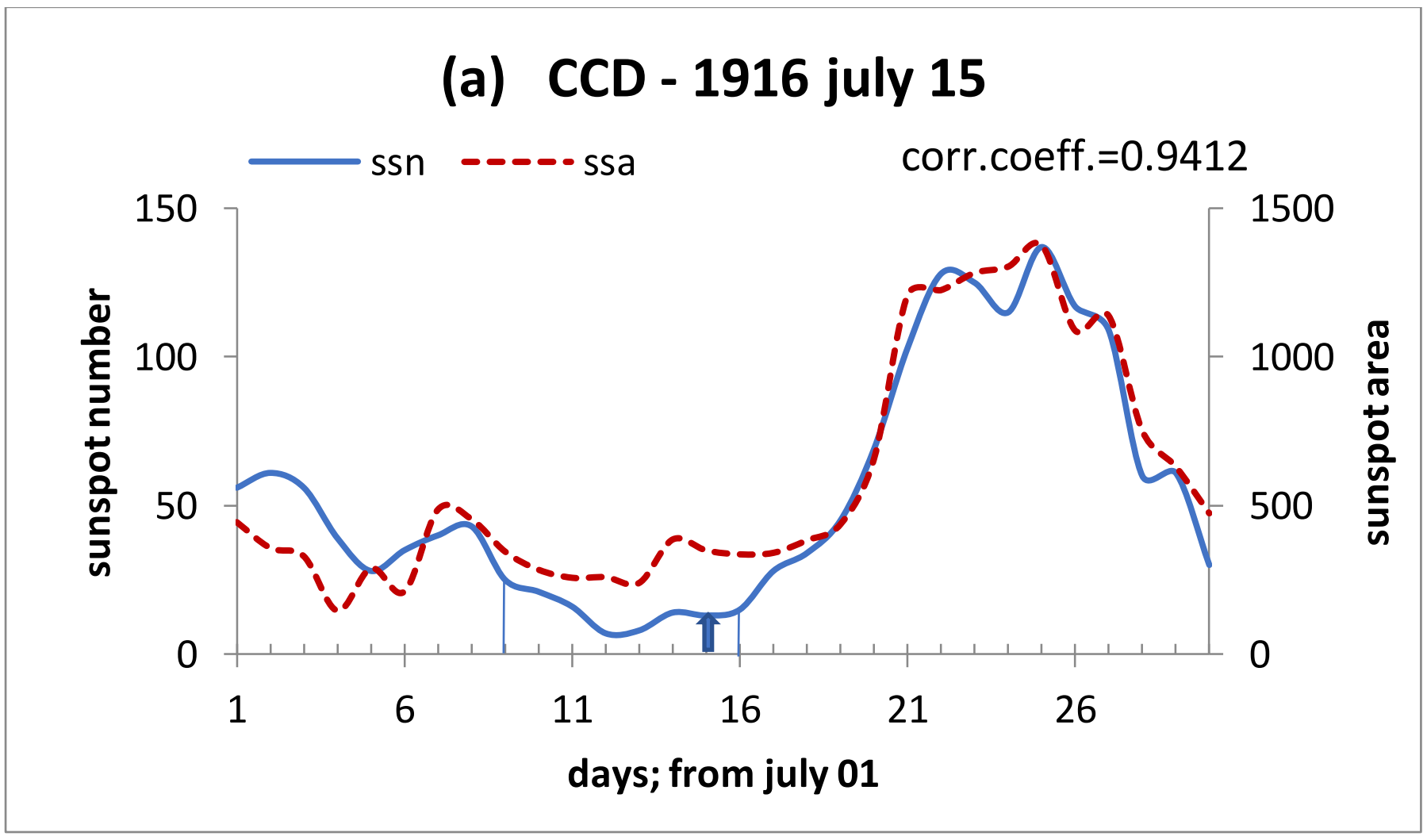

**Fig 1 (a)**

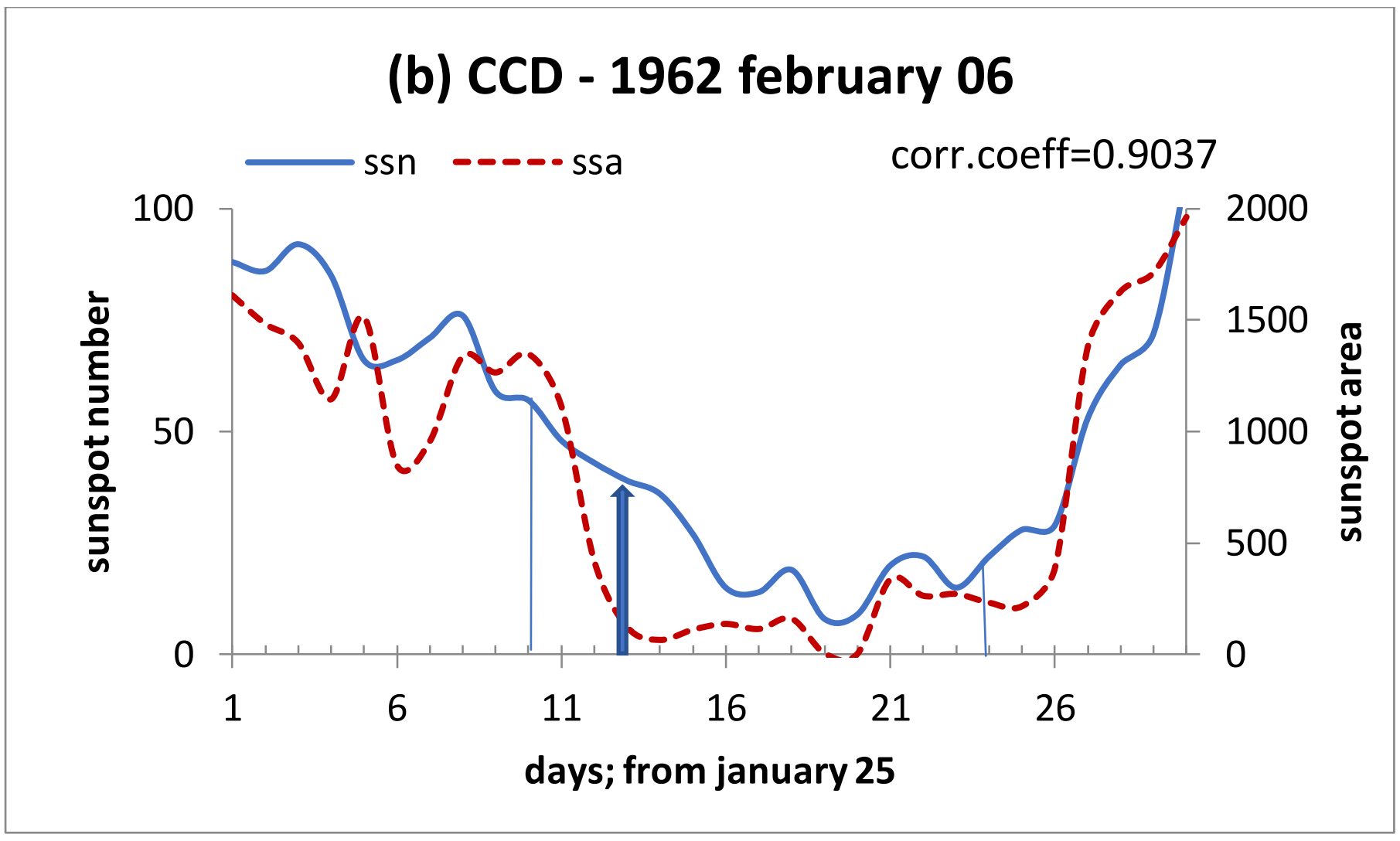

**Fig 1 (b)**

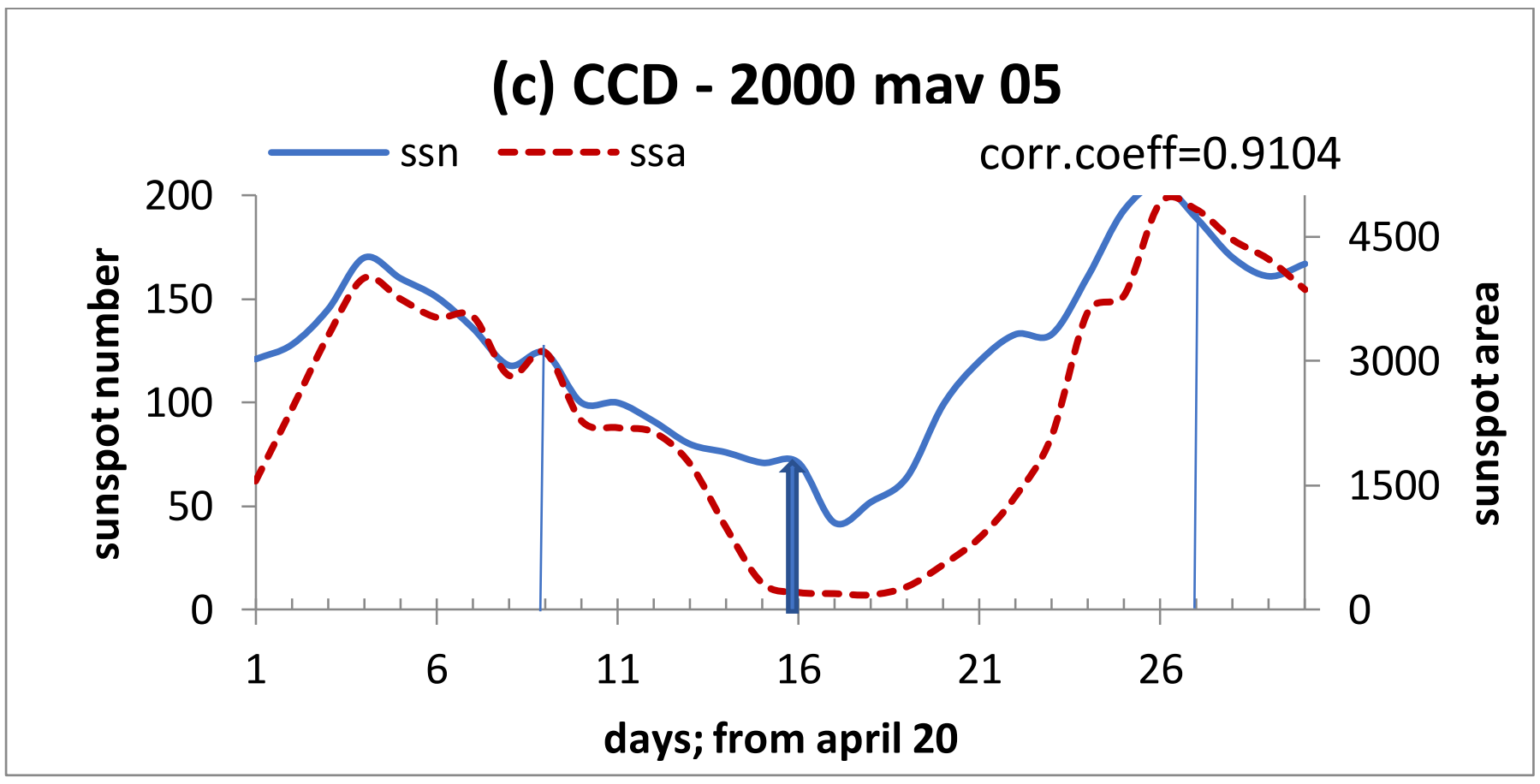

**Fig 1(c)**

**Figure 1 Variation of daily international sunspot number (R) given as solid curved line and Greenwich daily sunspot area (in units of MSH) given as dotted curved lines during selected planetary conjunction periods: (a) 1916 July (b) 1962 February and (c) 2000 May. The thin vertical lines shows the beginning and end of conjunction days and the thick vertical arrow shows closest conjunction date (CCD).**

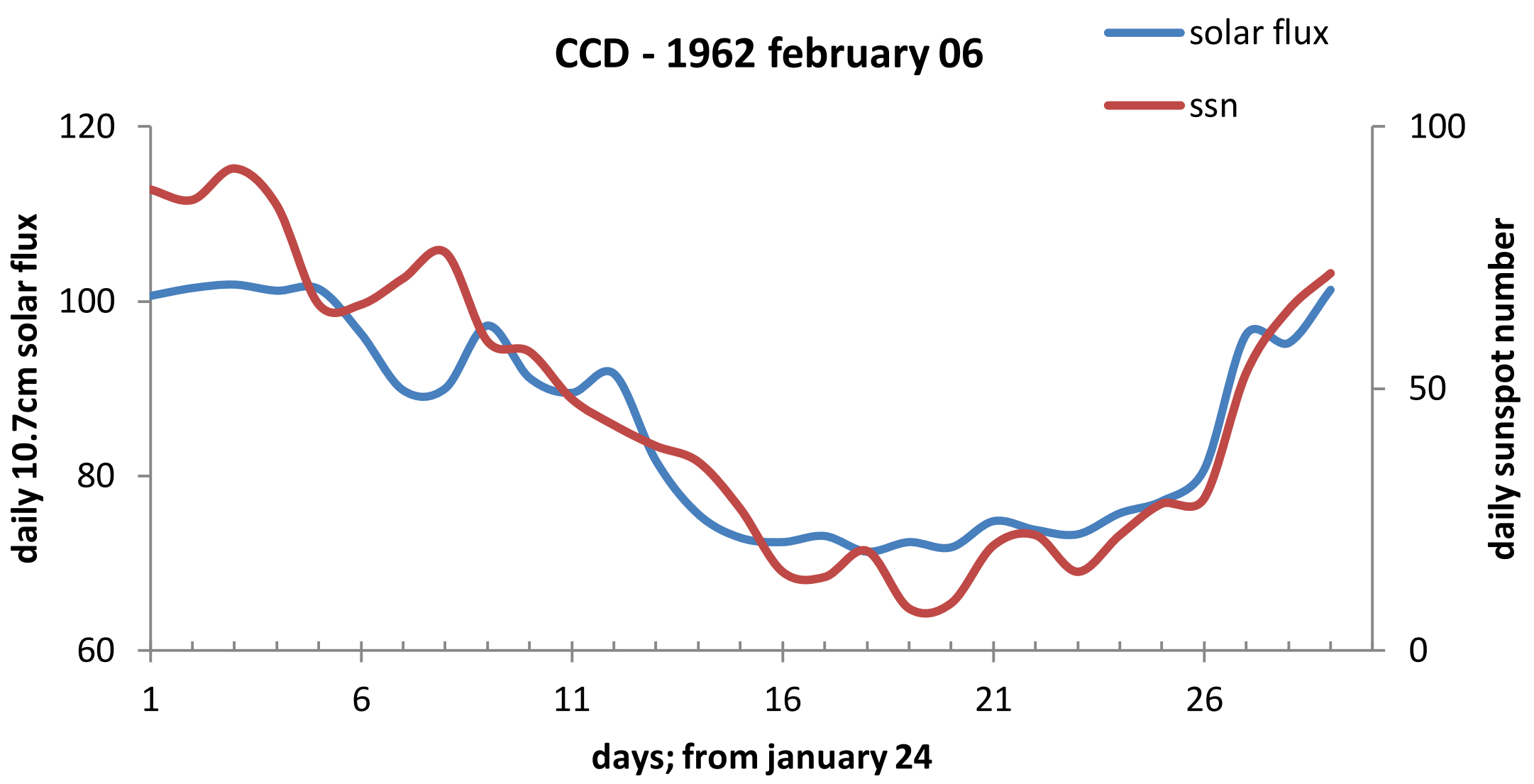

**Fig 2 Solar daily 10.7 cm radio flux variations during the 1962 February planetary conjunction period. The solar 10.7 cm flux changes are correlated with sunspot number ( ssn) variations during this MPC .**

## 3.2 On the nature of sunspot activity decreases during MPC

We could find 93 MPC with associated sunspot activity decreases out of 128 MPC studied for the years 1850-2000 AD. We classified the variations in daily sunspot number in connection with the planetary conjunctions into different categories based on the following scheme.

(a) **CC** - the daily sunspot number shows a characteristic decrease in a narrow region around the date of closest conjunction. (Examples: In Figure 2.1 CCD – 1863 October 06, CCD – 1921 September 05, CCD – 1998 may 28 etc.)

(b) **EC** – the daily sunspot number shows a notable decrease during the entire duration of planetary conjunction period. (Examples: In Figure 2.1 CCD – 1856 June 23, CCD - 1916 July 15, CCD- 1976 July 25 etc.)

(c) **CCPE** – the daily sunspot number shows a notable decrease only during the early part of the conjunction period including the date of CC. (Examples: In Figure 2.1 CCD – 1906 February 20, CCD – 1922 October 10, CCD – 1985 December 01 etc.)

(d) **CCPA** – the daily sunspot number shows a notable decrease only during the later part of the conjunction period including the date of closest conjunction. (Examples: In Figure 2.1 CCD – 1857 April 17, CCD – 1910 October 31, CCD – 1955 August 17 etc.)

(e) **NE** – there is no notable change in the daily sunspot variation during the conjunction period. (Examples: In Figure 2.1 CCD – 1870 December 08, CCD – 1938 February 17, CCD – 1950 January 22 etc.)

(f) **PE** - the daily sunspot number shows a decrease only during the early part of the conjunction period excluding closest conjunction.(Examples: In Figure 2.1 CCD – 1875 October 26, 1920 September 04 and CCD – 1946 October 15 etc.)

(g) **PA** – the daily sunspot number shows a decrease only during the later part of the conjunction period excluding closest conjunction. (Examples: In Figure 2.1 CCD – 1858 May 01, CCD- 1934 October 30 and CCD – 1946 November 19)

Examples are given in Figure 3.

For each of these MPC ( except for NE cases) we have determined the following parameters

If $R_{max}$ is the maximum daily sunspot number just before MPC and $R_{min}$ is the minimum daily sunspot number during the MPC period then we define $\delta_R$ as

$$\delta_R = \frac{R_{max} - R_{min}}{R_{max} + R_{min}} \qquad (1)$$

If $A_{max}$ is the maximum daily sunspot area just before MPC and $A_{min}$ is the maximum daily sunspot area durig the MPC period then we define $\delta_A$ as

$$\delta_A = \frac{A_{Max} - A_{Min}}{A_{Max} + A_{Min}} \qquad (2)$$

If $S_{max}$ is the maximum daily 10.7 cm solar radio flux just before MPC and $S_{min}$ is the minimum daily 10.7 cm solar radio flux during the MPC period then we define $\delta_s$ as

$$\delta_s = \frac{S_{max} - S_{min}}{S_{max} + S_{min}} \qquad (3)$$

Duration of sunspot activity decrease in days during the MPC period

For each of these 93 MPC we have calculated the above parameter and the results are given in Table 1 where we have also given the classification of sunspot activity decrease ( CCPA etc) , close conjunction date, planets in conjunction, maximum angular separation in geocentric longitude between the planets in conjunction. The details of δR calculations and duration of sunspot activity decrease during the MPC periods is given in Table 2.

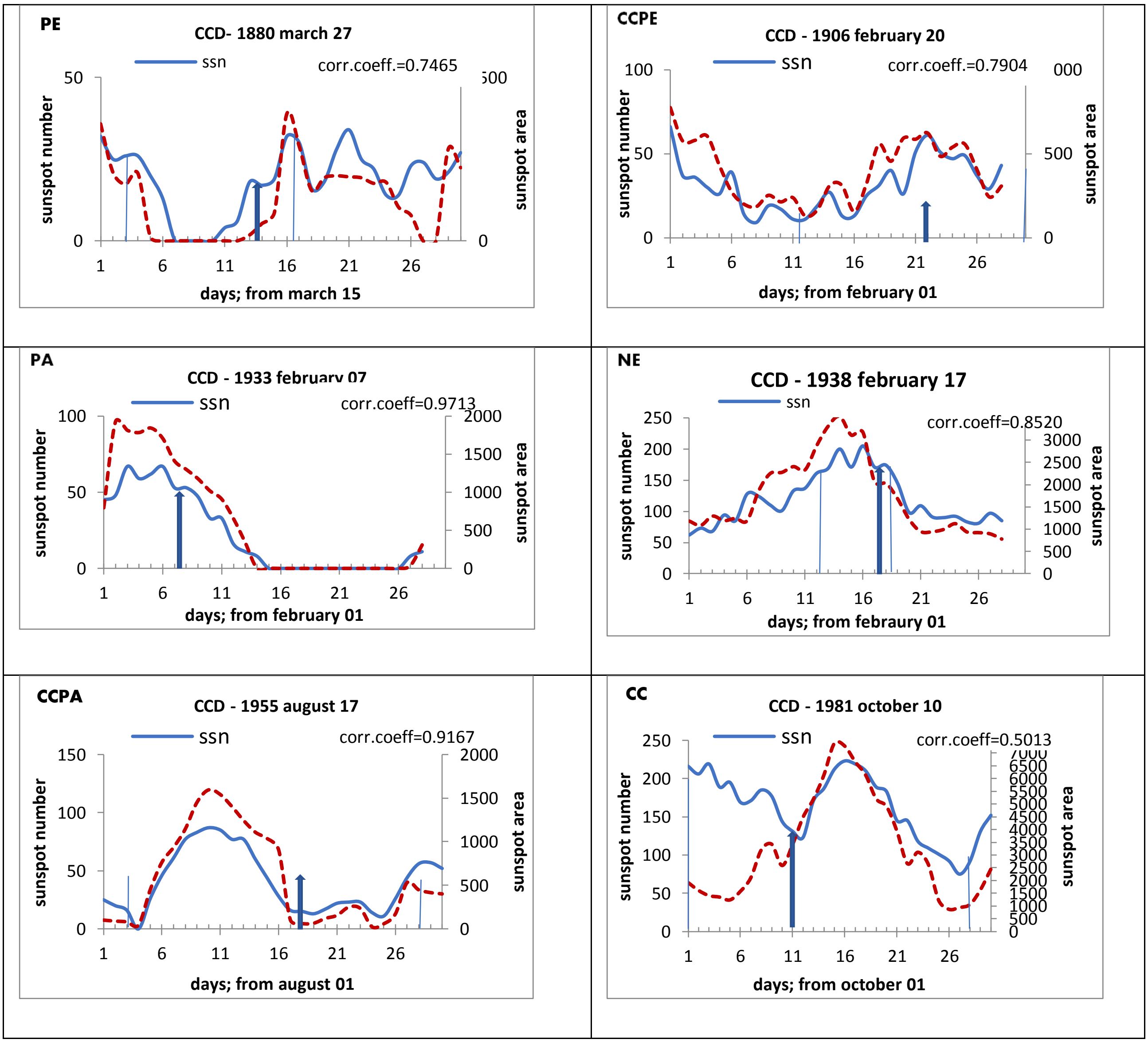

**Figure 3: Variation of daily international sunspot number (R) given as solid curved line and Greenwich daily sunspot area (in units of MSH) given as dotted curved lines during selected planetary conjunction periods for the category of PE, CCPE, PA, NE, CCPA & CC: The thin vertical lines shows the beginning and end of conjunction days and the thick vertical arrow shows closest conjunction date (CCD).**

The duration of sunspot decrease during different MPC studied during the years 1850-2000 AD is found to vary between 3-22 days . The frequency of occurrences of different categories of sunspot decreases during MPC studied for the period 1850-2000 AD is given in Table 3.

## 4. Extreme Space weather events and Multiple planetary conjunction periods during 1850-2000 AD

Weak solar activity conditions are expected to be associated with calm space weather conditions even though there can be few exceptions. In the previous section it is found that there is a high probability of low sunspot activity during multiple planetary conjunction periods. In this section we will study whether occurrences of severe space weather events such as intense geomagnetic storms,intense solar flares and 30 MeV solar proton events happen during multiple planetary conjunction periods during 1850 -2000 AD. The extreme space weather event data is obtained from published literature ( Cliver and Svaalgard,2004). A list of all such severe space weather events during this long period is given in Table 4 . None of these severe space weather events are found to occur during MPC periods which is also shown in this Table.108 out of 109 extreme space weather events are not found to occur during the MPC periods.

## 5 Space missions and Multiple planetary conjunctions

Understanding space weather conditions is important for the operations of technological systems in space like satellites . It also helps to plan space missions and related space craft launches from Earth. Calm solar activity conditions are understood to be generally associated with calm space weather conditions even though there are exceptions.

In Table 5 and Table 6 we have given list of lunar and human space flight missions associated with MPC's during the years 1957-2000 AD. Out of 22 space missions associated with planetary conjunction periods only 4 failed giving a success rate of more than 80 %.

## 6 Discussion

From our detailed investigations using relevant data for more than 150 years ( 1850-2000 AD) we could find that multiple planetary conjunction periods are associated with notable decreases in sunspot activity and relatively calm space weather conditions. We can use these results for solar-terrestrial prediction purposes and space mission planning. It can inspire further studies to understand its possible physical causes. This study stands apart from the previous studies on the effect of planetary dynamics on the solar activity variability.

Sunspot activity decrease is observed for 70% of multiple planetary conjunctions studied during 1850-2000.The duration of this decrease range from few days to more than two weeks. The sunspot activity decrease is observed for MPC both during sunspot minima and maxima phases. However the magnitude of decrease shows a weak dependence on the background level of daily sunspot activity during the conjunction period. In Figure 4 we have plotted Rmax against δR . The linear regression best fit line in this plot suggest an inverse relation between two parameters inspite of the large scatter in the data points . δR becomes 100% when $R_{min}$ during a MPC is zero which happens generally during low sunspot activity periods. When we compare δR, δA and δSduring different MPC as shown in Table we find δA is highest followed by δR and δS is the least of the three. As a typical example during the close MPC which occurred during May 2000 δR is found to be 49.4 % δA is 83.4 % and δS is only 23.8 %. Generally δS values are found to be only one-third or one fourth of δA. values for typical MPC. So among the sunspot activity related parameters sunspot area shows significant decreases during the the planetary conjunction periods while solar 10.7 cm flux changes are observed to be relatively small.

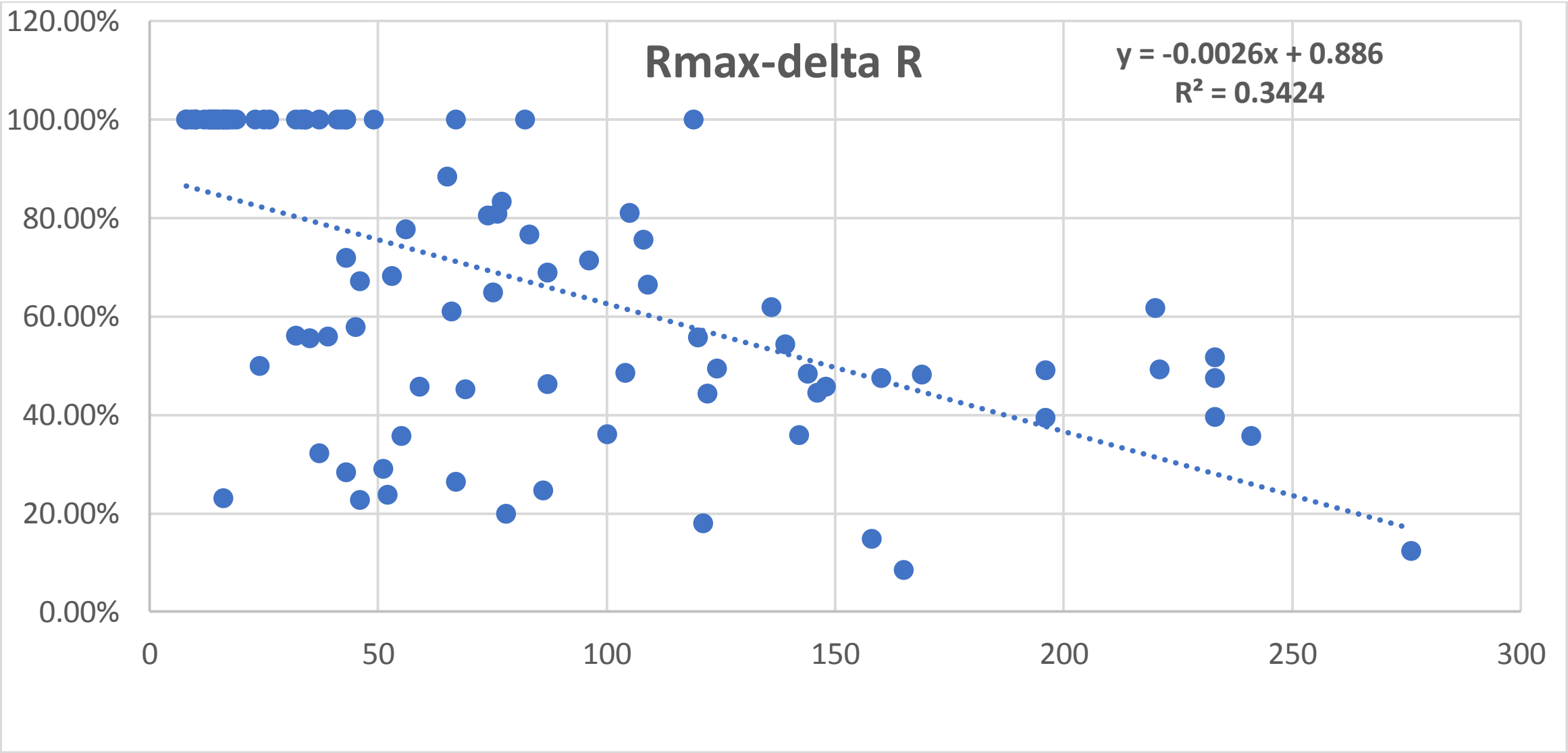

**Fig 4 : Linar fit between maximum sunspot number prior to MPC and percentage of sunspot number decreases during MPC for selected MPC between 1850-2000 AD**

We have found the absence of extreme space weather events during the almost all multiple planetary conjunction periods during the years 1850-2000 AD. This includes occurrences of very intense geomagnetic storms, 30 MeV solar proton events and intense solar flares. which belong the category of extreme space weather activity near Earth. From an earlier study it is found that only weak to moderate intensity geomagnetic storms are likely to occur during MPC periods Some isolated 10 MeV proton events have occurred during our multiple planetary conjunction periods.

We have studied the association between Lunar and human space missions during 1957-2000 and the multiple planetary conjunctions. Many of these missions which are carried out during planetary conjunction periods are found to be successful. We do not assign high statistical significance for these results. Since low sunspot activity and calm space weather conditions are most probable during planetary conjunctions for future multiple planetary conjunction periods they can be considered for of short space missions in future.

Results of the present study points out the possibilities of inferring solar activity and space weather conditions in the past MPC periods and predicting the same for future MPC periods. Mees paper gives the list of all MPC's back to 3102 BC similar to the recent 1962 Feb and 2000 May close MPC events studied in this paper. The dates of all close triple planetary conjunctions during 1971-2056 AD is now available (http://stevealbers.net/albers/ast/conjun/conjun.html ). We have studied the solar/sunspot activity decreases , occurences of space weather events and human space missions during these triple planetary conjunction periods between 2002-2016 AD. The results are given in Table 7 which are almost similar to our results from our studies during 1850-2000 AD. This will provide an independent conformation of our conclusions.

## 7 Conclusions

(i) Significant decreases in daily sunspot activity is found for majority of multiple planetary conjunction periods during 1850-2000 AD. Similar results is also found for daily solar 10.7 cm radio flux during 1958-2000 AD. The above effect is found to be independent of the phase of sunspot cycle

(ii) Occurrences of extreme space weather events are found to be very rare during multiple planetary conjunction periods as evident from analysis of relevant data for the years 1850 – 2000 AD.

(iii) Lunar and human space flight missions during the space craft era are found to be generally successful during planetary conjunction periods.

(iv) Future planetary conjunction periods we can plan safer short term space missions since it may be possible to predict solar-terrestrial conditions during such periods.

## References

Abreau J A, Beer J, Ferriz-Mas A, McCracken K G and Steinhilber F, Is there a planetary Influence on solar activity? , *Astron. Astrophys*, (2012), 548, A 88.

Ambroz P, Planetary influences on the large - scale distribution of solar activity, Solar Phys. 19, (1971), 480 – 482.

Chambers D.F, A Handbook of Descriptive and Practical Astronomy-Part 1, Clarendon Press, Oxford ( 1889), 68-71.

Charbonneau P, The rise and fall of the 1st solar cycle model, *J.Hist. Astro.,* 33, 4(113), (2002), 351-372.

Charbonneau P, The planetary hypothesis revived, *Nature*, 493, (2012), 613-614.

Cliver E W and Svalgaard L, The 1859 Solar terrestrial disturbance and the current limits of extreme space weather activity, *Solar Phys.*, 224, (2004), 407 – 422.

De La Rue W, Stewart B and Loewy B, Further investigations on planetary influence upon solar activity, *Proc. Royal Soc. London*, 20, (1872), 210 – 218.

Girish T E and Nisha N G, Planetary conjunctions, decrease of Solar activity and prediction of Space weather, Paper presented in 39th COSPAR Scientific Assembly 2012, Mysore.

Hung C C, Apparent relations between solar activity and solar tides caused by the planets, NASA/TM, (2007), 214817.

Ian Edmonds, Planetary model of sunspot emergence: A spectral and autocorrelation Analysis ( 2016) arXiv:1610.02632[astro-ph.SR]

Jose P D, Sun's motion and sunspots, *Astron. J*, 70, Number 3, (1965), 193 -200.

Meis S D and Meeus J, Quintuple planetary groupings – Rarity, historical events and popular beliefs, *J. Brit. Astron. Assoc, 104, No.6*, (1994), 293 – 297.

Nisha N.G, Investigations on the possible associations of sunspot activity variations with Planetary conjunctions during 1850-2000 , Mphil Thesis, University Kerala,Trivandrum India (2003).

Nisha N.G, Some studies on Planetary dynamics and Sunspot activity variations, PhD Thesis, , University Kerala, Trivandrum,India (2015).

Okhlopkov V P, Cycles of solar activity and the configurations of the planets, Jl. Phys. Conference Series, 409, (2013), 012199.

Scafetta N, Does the Sun work as a nuclear fusion amplifier of planetary tidal forcing?, A proposal for physical mechanism based on the mass – luminosity relation, J. Atmos. Solar Terr. Phys., 81, (2012), 27 – 40.

Schuster A, The influence of planets on the formation of Sun-spots, *Proc. of Royal Soc. London*, 135A, (1911), 309 - 323.

Stefani, F., Giesecke, A. & Weier, T. A Model of a Tidally Synchronized Solar Dynamo. *Sol Phys* **294,** 60 (2019). https://doi.org/10.1007/s11207-019-1447-1

Swamikannu Pillai L D, The Indian calendar and Indian astrology, *Asian Educational Services*, New Delhi, (1985).

Wolff C L and Patrone P, A new way that planets can affect the Sun, Solar Phys., 266, (2010), 227 – 246

**Table 1 : Decrease in (%) of sunspot activity and solar 10.7 cm flux during multiple planetary conjunction periods during 1850-2000 AD**

| | | | | | | | |
|---|---|---|---|---|---|---|---|
| | | | | | | | |
| | | | | | | | |
| **Conjunction period** | **Planets** | **CC date** | **ΔΘ (deg)** | **Type of SSA decr** | **Δ R** | **Δ A** | **Δ S** |
| 1850 Aug 29 – Sep 18 | Sun, Mar, Mer, Jup | Sep-14 | 12.3 | CC | 24.64% | …….. | …….. |
| 1851 Oct 1 - Nov 08 | Sun, Mer, Jup, Ven | Nov-02 | 14.1 | PE | 48.57% | …….. | …….. |
| 1852 Nov 05 – Nov 19 | Sun, Mar, Mer, Jup | Nov-15 | 13.9 | CCPA | 81.03% | …….. | …….. |
| 1856 Jun 15 –Jun 30 | Sun, Mer, Ven, Sat | Jun-23 | 18.2 | EC | 100.00% | …….. | …….. |
| 1857 Apr 11 –Apr 21 | Sun, Mar, Mer, Jup | Apr-17 | 19.7 | CCPA | 100.00% | …….. | …….. |
| 1858 Apr 11 – May 11 | Sun, Mer, Jup, Ven | May-01 | 7.3 | PA | 80.49% | …….. | …….. |
| 1859 Jun 13 – Jul 11 | Sun, Mar, Mer, Jup | Jun-21 | 11.7 | PE | 44.38% | …….. | …….. |
| 1860 Jun30 - Jul 20 | Sun, Mer, Jup, Ven | Jul-05 | 15.2 | CCPA | 61.76% | …….. | …….. |
| 1861 Aug 29 – Sep 18 | Sun, Mar, Mer, Jup, Sat | Sep-03 | 10.5 | CCPA | 54.44% | …….. | …….. |
| 1862 Aug 25 - Sep 18 | Sun, Mer, Jup, Sat | Sep-09 | 7.2 | EC | 71.43% | …….. | …….. |
| 1863 Aug 30 - Sep 09 | Sun, Mar, Mer,Ven, Sat | Sep-04 | 18.3 | EC | 100.00% | …….. | …….. |
| 1863 sep 29 – Oct 16 | Sun, Mar, Mer, Sat | Oct-06 | 10.9 | CC | 64.84% | …….. | …….. |
| 1864 Sep 05 – Sep 21 | Sun, Mer, Ven, Sat | Sep-19 | 14.7 | PE | 68.25% | …….. | …….. |
| 1865 Oct 18 - Nov 06 | Sun, Mar, Mer, Sat | Oct-25 | 6.6 | EC | 100.00% | …….. | …….. |
| 1867 Oct 19 – Oct 24 | Sun, Mar, Mer, Sat | Oct-24 | 14.8 | EC | 100.00% | …….. | …….. |
| 1869 Apr 20 - May 01 | Sun, Mar, Mer, Jup | Apr-23 | 14.3 | CCPA | 60.98% | …….. | …….. |
| 1872 Jul 10 – Jul 20 | Sun, Mer, Jup, Ven | Jul-11 | 21.5 | PA | 49.04% | …….. | …….. |
| 1874 Jan 16 – Feb 05 | Sun, Mer, Ven, Sat | Jan-27 | 12 | CC | 66.41% | …….. | …….. |
| 1875 Oct 09 – Oct 30 | Sun, Mer, Jup, Ven | Oct-26 | 13.8 | PE | 100.00% | 100.00% | …….. |
| 1876 Jan 04 – Jan 19 | Sun, Mer, Ven, Sat | Jan-08 | 16.7 | EC | 100.00% | 100.00% | …….. |
| 1877 Mar 15 – Mar 17 | Sun, Mer, Ven, Sat | Mar-17 | 19.3 | EC | 100.00% | 100.00% | …….. |
| 1880 Mar 17-Mar 30 | Sun, Mer, Jup, Sat | Mar-27 | 6.3 | CCPE | 100.00% | 100.00% | …….. |
| 1881 Apr 29 - May 05 | Sun, Jup, Ven, Sat | May-01 | 16.6 | EC | 88.41% | 93.09% | …….. |
| 1882 Apr 19 –May 15 | Sun, Mer, Jup, Ven, Sat | May-10 | 18.6 | EC | 14.91% | 19.90% | …….. |
| 1888 Aug 02 – Aug 19 | Sun, Mer, Ven, Sat | Aug-13 | 18.1 | EC | 28.36% | 24.38% | …….. |
| 1891 Sep 04 – Sep 19 | Sun, Mar, Mer, Ven | Sep-12 | 16.9 | CC | 55.84% | 65.41% | …….. |
| 1893 Sep 30 | Sun, Mar, Mer, Sar | Sep-30 | 16.2 | PA | 36.05% | 43.73% | …….. |
| 1895 Oct 25 – Nov 18 | Sun, Mar, Mer, Sat | Oct-29 | 11.1 | PE | 100.00% | 99.47% | …….. |
| 1896 Jul 29 - Aug 15 | Sun, Mer, Jup, Ven | Aug-07 | 5.4 | EC | 100.00% | 100.00% | …….. |
| 1897 Sep 07 - Sep 15 | Sun, Mar, Mer, Jup | Sep-10 | 21.4 | CCPA | 76.60% | 93.27% | …….. |
| 1897 Nov 05 – Nov 29 | Sun, Mar, Mer, Sat | Nov-17 | 5 | EC | 100.00% | 100.00% | …….. |
| 1898 Nov 15 – Nov 28 | Sun, Mer, Ven, Sat | Nov-19 | 15.5 | EC | 55.56% | 88.00% | …….. |
| 1899 Oct 17 - Nov 08 | Sun, Mer, Jup, Ven | Oct-19 | 16.5 | EC | 100.00% | 100.00% | …….. |
| 1900 Dec27– '01Jan 05 | Sun, Mer, Jup, Sat | Dec-30 | 17.2 | EC | 100.00% | | …….. |
| 1902 Jan 01 - Jan 09 | Sun, Mar, Mer, Jup | Jan-06 | 17.3 | PE | 100.00% | 100.00% | …….. |
| 1903 Jan 01 - 18 | Sun, Mer, Ven, Sat | Jan-05 | 11.8 | PA | 100.00% | 100.00% | …….. |
| 1903 Jan 20 - Feb 02 | Sun, Mer, Jep, Ven | Jan-29 | 12.2 | PE | 100.00% | 100.00% | …….. |
| 1905 Apr 11 - May 05 | Sun, Mer, Jup, Ven | Apr-30 | 13.4 | PE | 45.68% | 55.09% | …….. |
| 1906 Feb 10 - 28 | Sun, Mer, Ven, Sat | Fen 20 | 10 | CCPA | 56.00% | 32.28% | …….. |
| 1906 May 03 -13 | Sun, Mar, Jup, Ven | May-08 | 15.5 | CCPE | 23.81% | 47.63% | …….. |
| 1906 Jun 02-17 | Sun, Mar, Mer, Jup | Jun-12 | 12.6 | CCPA | 75.61% | 71.91% | …….. |
| 1907 Jul 25 - Aug 06 | Sun, Mer, Jup, Ven | Aug-01 | 15.4 | CCPA | 26.42% | 52.33% | …….. |
| 1908 Aug 07 - 30 | Sun, Mer, Mar, Jup | Aug-18 | 4.8 | EC | 61.90% | 84.55% | …….. |
| 1910 Oct 10 - Nov 09 | Sun, Mars, Mer, Jup, Ven | Oct-31 | 12 | CCPA | 100.00% | 100.00% | …….. |
| 1912 May 27 - Jun4 | Sun, Mer, Ven, Sat | Jun-01 | 16.4 | PA | 100.00% | 100.00% | …….. |
| 1914 Jan 07 - 30 | Sun, Mer, Jup, Ven | Jan-23 | 7.1 | EC | 100.00% | 100.00% | …….. |
| 1915 Mar 04 - 30 | Sun, Mar, Mer, Jup | Mar-04 | 20.2 | PA | 83.33% | 79.76% | …….. |
| 1916 Jul 09 - Jul 16 | Sun, Mer, Jup, Ven | Jul-15 | 14 | EC | 72.00% | 33.98% | …….. |
| 1917 Apr 13 - May 16 | Sun, Mer, Jup, Ven | Apr 13, May 13 | 18.7 | CCPE | 47.47% | 49.84% | …….. |
| 1917 Jul 10 – Jul 22 | Sun, Mer, Ven, Sat | Jul-16 | 17 | CCPA | 45.81% | 58.99% | …….. |
| 1920 Jul 20 – Aug 25 | Sun, Mer, Jup, Ven | Aug-20 | 19.8 | CC | 67.27% | 39.01% | …….. |
| 1921 Sep 01 – Sep 21 | Sun, Mer, Ven, Sat | Sep-05 | 11.1 | CC | 100.00% | 100.00% | …….. |
| 1922 Sep 29 - Oct 24 | Sun, Mer, Jup, Sat | Oct-10 | 7 | CCPA | 100.00% | 100.00% | …….. |
| 1923 Nov 17 - Nov 23 | Sun, Mer, Jup, Ven | Nov-19 | 16.5 | EC | 100.00% | 100.00% | …….. |
| 1929 Dec 01 – Dec 22 | Sun, Mar, Mer, Sat | Dec-15 | 17.8 | CCPE | 44.55% | 52.12% | …….. |
| 1930 May 10 – May 17 | Sun, Mer, Jup, Ven | May-13 | 17.5 | CCPE | 32.14% | 59.33% | …….. |
| 1933 Feb 07 | Sun, Mer, Ven, Sat | Feb-07 | 20.8 | PA | 100.00% | 100.00% | …….. |
| 1934 Jan 20 - Feb 12 | Sun, Mer, Ven, Sat | Jan-23 | 15.4 | EC | 100.00% | 100.00% | …….. |
| 1934 Oct 30 - Nov 16 | Sun, Mer, Jup, Ven | Oct-30 | 10 | PA | 100.00% | 100.00% | …….. |
| 1935 Jan 22 - Feb 17 | Sun, Mer, Ven, Sat | Jan-29 | 13.7 | CC | 56.10% | 93.56% | …….. |
| 1938 Mar 06 – Mar 30 | Sun, Mer, Ven, Sat | Mar-20 | 7.1 | CCPA | 20.00% | 25.92% | …….. |
| 1941 Apr 29 - May 18 | Sun, Mer, Jup, Ven, Sat | May-07 | 7 | CCPA | 57.89% | 88.03% | …….. |
| 1944 Jun 16 - Jul 09 | Sun, Mer, Ven, Sat | Jun-28 | 7.7 | PA | 100.00% | 100.00% | …….. |
| 1944 Aug 21 - Aug 31 | Sun, Mer, Jup, Ven | Aug-26 | 12.6 | PE | 100.00% | 100.00% | …….. |
| 1946 Oct 09 – Oct 20 | Sun, Mar, Mer, Jup | Oct-15 | 10 | PE | 48.45% | 28.25% | …….. |
| 1946 Nov 17 – Nov 21 | Sun, Mer, Jup, Ven | Nov-19 | 18.4 | PA | 48.25% | 19.42% | …….. |
| 1947 Oct28 - Nov 05 | Sun, Mer, Jup, Ven | Nov-03 | 13.8 | EC | 51.79% | 59.73% | …….. |
| 1948 Dec 15– ‘49Jan06 | Sun, Mar, Mer, Jup | Dec-28 | 13.9 | CCPA | 49.32% | 48.74% | |
| 1950 Sep 20 – Sep 30 | Sun, Mer, Ven, Sat | Sep-25 | | EC | 46.22% | 73.92% | |
| 1951 Mar 08 – Mar 16 | Sun, Mar, Mer, Jup | Mar-10 | 17.2 | EC | 35.80% | 23.68% | |
| 1954 Nov 08 – Nov 23 | Sun, Mer, Ven, Sat | Nov-21 | 17.4 | EC | 100.00% | 100.00% | |
| 1955 Aug 17 – Aug 27 | Sun, Mar, Jup, Ven | Aug-17 | 18.3 | CCPA | 68.93% | 95.04% | |
| 1957 Sep 30 – Oct 24 | Sun, Mar, Mer, Jup | Oct-01 | 14 | CCPA | 12.42% | 11.08% | |
| 1958 Oct 18 – Oct 30 | Sun, Mer, Jup, Ven | Oct-24 | 19.5 | CC | 35.77% | 61.31% | 11.86% |
| 1962 Feb 03 – Feb 17 | Sun, Mar, Mer, Jup, Ven, Sat | Feb-25 | 17 | EC | 80.95% | 99.41% | 20.74% |
| 1964 Apr 30 - May 03 | Sun, Mar, Mer, Jup | Apr-30 | 17.9 | EC | 100.00% | 29.51% | 25.73% |
| 1965 Feb 22 – Mar 07 | Sun, Mer, Ven, Sat | Mar-05 | 20.3 | CCPE | 100.00% | 100.00% | 2.57% |
| 1966 Feb 11 - Mar 14 | Sun, Mar, Mer, Sat | Feb-25 | 17.1 | CCPA | 100.00% | 100.00% | 4.18% |
| 1970 Oct 25 - Nov 09 | Sun, Mer, Jup, Ven | Nov-04 | 12.7 | CC | 35.89% | 76.68% | 6.98% |
| 1971 Nov 08 – Nov 29 | Sun, Mer, Jup, Ven | Nov-11 | 0.8 | CCPA | 45.26% | 50.53% | 17.79% |
| 1973 May 21 - Jun12 | Sun, Mer, Ven, Sat | May-30 | 17.4 | CCPA | 77.78% | 100.00% | 12.03% |
| 1976 Jul 12 - Aug 04 | Sun, Mer, Ven, Sat | Jul-25 | 9.6 | EC | 100.00% | 100.00% | 12.75% |
| 1979 Aug 02 – Aug 14 | Sun, Mer, Jup, Ven | Aug-06 | 5.6 | CCPE | 18.05% | 22.84% | 10.33% |
| 1979 Sep 01 – Sep 17 | Sun, Mer, Ven, Sat | Sep-13 | 9.8 | EC | 8.55% | 38.10% | 18.10% |
| 1980 Aug 30 - Sep 17 | Sun, Mer, Jup, Sat | Sep-10 | 6.2 | CCPE | 47.47% | 60.06% | 14.68% |
| 1981 Oct 01 – Oct 25 | Sun, Mer, Jup, Sat | Oct-10 | 11.5 | CC | 39.52% | 24.68% | 20.30% |
| 1985 Nov 22 – Dec 09 | Sun, Mer, Jup, Ven | Dec-01 | 15.3 | CCPE | 100.00% | 47.44% | 24.89% |
| 1989 May 06 – May 25 | Sun, Mer, Jup, Ven | May-22 | 19.4 | CC | 39.50% | 30.97% | 9.48% |
| 1994 Feb 07 – Feb 17 | Sun, Mer, Ven, Sat | Feb-16 | 17.5 | CC | 22.67% | 51.43% | 10.11% |
| 1995 Nov 30 – Dec 06 | Sun, Mar, Mer, Jup, Ven | Dec-01 | 6.4 | CCPA | 23.08% | 20.00% | 7.49% |
| 1997 Mar 12 – Mar 29 | Sun, Mer, Ven, Sat | Mar-21 | 20.1 | CCPA | 100.00% | 100.00% | 4.90% |
| 1998 May 23 - Jun 01 | Sun, Mar, Mer, Sat | May-28 | 12.1 | CC | 29.11% | 86.44% | 25.4% |
| 2000 Apr 28 – May 16 | Sun, Mar, Mer, Jup, Sat | May-05 | 20.1 | EC | 49.40% | 84.34% | 23.8 % |
| | | | | | | | |

**Table 2. Calculations of normalized sunspot activity decreases during multiple planetary conjunction periods of the years 1850-2000 AD**

| Conjunction date | Duration of SSN decrease | $R_{max}$ | $R_{min}$ | $\delta_R = \frac{R_{max} - R_{min}}{R_{max} + R_{min}}$ |
|---|---|---|---|---|
| 1850 Aug 29 – Sep 18 | 3 | 86 | 52 | 24.64% |
| 1851 Oct 1 - Nov 08 | 8 | 104 | 36 | 48.57% |
| 1852 Nov 05 – Nov 19 | 4 | 105 | 11 | 81.03% |
| 1856 Jun 15 –Jun 30 | 16 | 15 | 0 | 100.00% |
| 1857 Apr 11 –Apr 21 | 6 | 32 | 0 | 100.00% |
| 1858 Apr 11 – May 11 | 8 | 74 | 8 | 80.49% |
| 1859 Jun 13 – Jul 11 | 7 | 122 | 47 | 44.38% |
| 1860 Jun30 - Jul 20 | 14 | 220 | 52 | 61.76% |
| 1861 Aug 29 – Sep 18 | 15 | 139 | 41 | 54.44% |
| 1862 Aug 25 - Sep 18 | 25 | 96 | 16 | 71.43% |
| 1863 Aug 30 - Sep 09 | 11 | 34 | 0 | 100.00% |
| 1863 sep 29 – Oct 16 | 5 | 75 | 16 | 64.84% |
| 1864 Sep 05 – Sep 21 | 3 | 53 | 10 | 68.25% |
| 1865 Oct 18 - Nov 06 | 20 | 42 | 0 | 100.00% |
| 1867 Oct 19 – Oct 24 | 6 | 18 | 0 | 100.00% |
| 1869 Apr 20 - May 01 | 0 | 66 | 16 | 60.98% |
| 1872 Jul 10 – Jul 20 | 6 | 196 | 67 | 49.04% |
| 1874 Jan 16 – Feb 05 | 7 | 109 | 22 | 66.41% |
| 1875 Oct 09 – Oct 30 | 17 | 37 | 0 | 100.00% |
| 1876 Jan 04 – Jan 19 | 15 | 8 | 0 | 100.00% |
| 1877 Mar 15 – Mar 17 | 8 | 14 | 0 | 100.00% |
| 1880 Mar 17-Mar 30 | 8 | 49 | 0 | 100.00% |
| 1881 Apr 29 - May 05 | 5 | 65 | 4 | 88.41% |
| 1882 Apr 19 –May 15 | 7 | 158 | 117 | 14.91% |
| 1888 Aug 02 – Aug 19 | 10 | 43 | 24 | 28.36% |
| 1891 Sep 04 – Sep 19 | 6 | 120 | 34 | 55.84% |
| 1893 Sep 30 | 5 | 100 | 47 | 36.05% |
| 1895 Oct 25 – Nov 18 | 3 | 119 | 0 | 100.00% |
| 1896 Jul 29 - Aug 15 | 18 | 33 | 0 | 100.00% |
| 1897 Sep 07 - Sep 15 | 7 | 83 | 11 | 76.60% |
| 1897 Nov 05 – Nov 29 | 7 | 16 | 0 | 100.00% |
| 1898 Nov 15 – Nov 28 | 14 | 35 | 10 | 55.56% |
| 1899 Oct 17 - Nov 08 | 7 | 15 | 0 | 100.00% |
| 1900 Dec27–'01Jan 05 | 10 | 14 | 0 | 100.00% |
| 1902 Jan 01 - Jan 09 | 4 | 25 | 0 | 100.00% |
| 1903 Jan 01 - 18 | 10 | 16 | 0 | 100.00% |
| 1903 Jan 20 - Feb 02 | 3 | 13 | 0 | 100.00% |
| 1905 Apr 11 - May 05 | 7 | 59 | 22 | 45.68% |
| 1906 Feb 10 - 28 | 9 | 39 | 11 | 56.00% |
| 1906 May 03 -13 | 5 | 52 | 32 | 23.81% |
| 1906 Jun 02-17 | 6 | 108 | 15 | 75.61% |
| 1907 Jul 25 - Aug 06 | 5 | 67 | 39 | 26.42% |
| 1908 Aug 07 - 30 | 7 | 136 | 32 | 61.90% |
| 1910 Oct 10 - Nov 09 | 13 | 43 | 0 | 100.00% |
| 1912 May 27 - Jun4 | 2 | 13 | 0 | 100.00% |
| 1914 Jan 07 - 30 | 20 | 17 | 0 | 100.00% |
| 1915 Mar 04 - 30 | 5 | 77 | 7 | 83.33% |
| 1916 Jul 09 - Jul 16 | 8 | 43 | 7 | 72.00% |
| 1917 Apr 13 - May 16 | 6 | 160 | 57 | 47.47% |
| 1917 Jul 10 – Jul 22 | 5 | 148 | 55 | 45.81% |
| 1920 Jul 20 – Aug 25 | 6 | 46 | 9 | 67.27% |
| 1921 Sep 01 – Sep 21 | 8 | 82 | 0 | 100.00% |
| 1922 Sep 29 - Oct 24 | 15 | 10 | 0 | 100.00% |
| 1923 Nov 17 - Nov 23 | 7 | 9 | 0 | 100.00% |
| 1929 Dec 01 – Dec 22 | 22 | 146 | 56 | 44.55% |
| 1930 May 10 – May 17 | 5 | 37 | 19 | 32.14% |
| 1933 Feb | 11 | 67 | 0 | 100.00% |
| 1934 Jan 30 - Feb 12 | 20 | 12 | 0 | 100.00% |
| 1934Oct 30 - Nov 16 | 6 | 16 | 0 | 100.00% |
| 1935 Jan 22 - Feb 17 | 7 | 32 | 9 | 56.10% |
| 1938 Mar 06 – Mar 30 | 2 | 78 | 52 | 20.00% |
| 1941 Apr 29 - May 18 | 5 | 45 | 12 | 57.89% |
| 1944 Jun 16 - Jul 09 | 5 | 8 | 0 | 100.00% |
| 1944 Aug 21 -Aug 31 | 4 | 34 | 0 | 100.00% |
| 1946 Oct 09 – Oct 20 | 3 | 144 | 50 | 48.45% |
| 1946 Nov 17 – Nov 21 | 2 | 169 | 59 | 48.25% |
| 1947 Oct28 - Nov 05 | 8 | 233 | 74 | 51.79% |
| 1948 Dec 15–‘49Jan06 | 10 | 221 | 75 | 49.32% |
| 1950 Sep 20 – Sep 30 | 11 | 87 | 32 | 46.22% |
| 1951 Mar 08 – Mar 16 | 8 | 55 | 26 | 35.80% |
| 1954 Nov 08 – Nov 23 | 14 | 41 | 0 | 100.00% |
| 1954 Nov 18 – Nov 23 | 6 | 24 | 8 | 50.00% |
| 1955 Aug 17 – Aug 27 | 7 | 87 | 16 | 68.93% |
| 1957 Sep 30 – Oct 24 | 2 | 276 | 215 | 12.42% |
| 1958 Oct 18 – Oct 30 | 6 | 241 | 114 | 35.77% |
| 1962 Feb 03 – Feb 17 | 15 | 76 | 8 | 80.95% |
| 1964 Apr 30 - May 03 | 4 | 23 | 0 | 100.00% |
| 1965 Feb 22 – Mar 07 | 4 | 26 | 0 | 100.00% |
| 1966 Feb 11 - Mar 14 | 9 | 17 | 0 | 100.00% |
| 1970 Oct 25 - Nov 09 | 7 | 142 | 67 | 35.89% |
| 1971 Nov 08 – Nov 29 | 16 | 69 | 26 | 45.26% |
| 1973 May 21 - Jun12 | 7 | 56 | 7 | 77.78% |
| 1976 Jul 12 - Aug 04 | 22 | 10 | 0 | 100.00% |
| 1979 Aug 02 – Aug 14 | 13 | 121 | 84 | 18.05% |
| 1979 Sep 01 – Sep 17 | 0 | 165 | 139 | 8.55% |
| 1980 Aug 30 - Sep 17 | 3 | 233 | 83 | 47.47% |
| 1981 Oct 01 – Oct 25 | 4 | 233 | 101 | 39.52% |
| 1985 Nov 22 – Dec 09 | 7 | 43 | 0 | 100.00% |
| 1989 May 06 – May 25 | 2 | 196 | 85 | 39.50% |
| 1994 Feb 07 – Feb 17 | 2 | 46 | 29 | 22.67% |
| 1995 Nov 30 – Dec 06 | 2 | 16 | 10 | 23.08% |
| 1997 Mar 12 – Mar 29 | 5 | 19 | 0 | 100.00% |
| 1998 May 23 - Jun 01 | 3 | 51 | 28 | 29.11% |
| 2000 Apr 28 – May 16 | 13 | 124 | 42 | 49.40% |

**Table 3: Frequency of occurrences of different categories of sunspot activity ( SSA) decreases during Planetary conjunction periods of 1850-2000 AD**

| Category of SSA decrease during MPC | No. of occurrences during 1850-2000 AD |
|---|---|
| CC | 13 |
| PE | 11 |
| CCPA | 21 |
| NE | 15 |
| EC | 29 |
| PA | 10 |
| CCPE | 11 |

**Table 4 Details of ExtremeSpace weather events between 1850-2000 AD. Association with MPC periods if any is indicated .**

| Sl. No. | Date of SW event | Type | Index of activity | Value | Association with MPC period if any |
|---|---|---|---|---|---|
| | 1851 August | >30 MeV SPE | Fluence (10^9 pr cm-2) | 9.3 | Nil |
| | 1859 August | >30 MeV SPE | Fluence (10^9 pr cm-2) | 18.8 | 1859 Aug 19- Aug 25(NE) |
| | 1859 September 01 | Solar Flare | Intensity | | Nil |
| | 1864 August | >30 MeV SPE | Fluence (10^9 pr cm-2) | 7 | Nil |
| | 1872 February 04 | GM Storm | Aa index in nT | 278 | Nil |
| | 1878 June | >30 MeV SPE | Fluence (10^9 pr cm-2) | 5 | Nil |
| | 1882 April 16 | GM Storm | Aa index in nT | 328 | Nil |
| | 1882 November 17 | GM Storm | Aa index in nT | 399 | Nil |
| | 1892 February 13 | GM Storm | Aa index in nT | 289 | Nil |
| | 1894 September | >30 MeV SPE | Fluence (10^9 pr cm-2) | 7.7 | Nil |
| | 1895 July | >30 MeV SPE | Fluence (10^9 pr cm-2) | 11.1 | Nil |
| | 1896 July | >30 MeV SPE | Fluence (10^9 pr cm-2) | 8 | Nil |
| | 1903 October 31 | GM Storm | Aa index in nT | 361 | Nil |
| | 1909 September 25 | GM Storm | Aa index in nT | 341 | Nil |
| | 1920 March 22 | GM Storm | Aa index in nT | 260 | Nil |
| | 1921 May 14 | GM Storm | Aa index in nT | 417 | Nil |
| | 1928 July 7 | GM Storm | Aa index in nT | 347 | Nil |
| | 1938 Jan 25 | GM Storm | Dst index in nT | -352 | Nil |
| | 1938 January 22 | GM Storm | Dst index in nT | -344 | Nil |
| | 1940 March 24 | GM Storm | Aa index in nT | 378 | Nil |
| | 1940 March 24 | GM Storm | Dst index in nT | -366 | Nil |
| | 1941 July 05 | GM Storm | Dst index in nT | -453 | Nil |
| | 1941 September 19 | GM Storm | Dst index in nT | -359 | Nil |
| | 1941 March01 | GM Storm | Dst index in nT | -382 | Nil |
| | 1942 February 28 | Solar Flare | Intensity | 3 | Nil |
| | 1946 February 07 | GM Storm | Aa index in nT | 254 | Nil |
| | 1946 July 25 | Solar Flare | Intensity | 3+ | Nil |
| | 1946 March 28 | GM Storm | Dst index in nT | -440 | Nil |
| | 1946 May | >30 MeV SPE | Fluence (10^9 pr cm-2) | 6 | Nil |
| | 1946 September 22 | GM Storm | Aa index in nT | 291 | Nil |
| | 1949 January 26 | GM Storm | Dst index in nT | -350 | Nil |
| | 1949 November 19 | Solar Flare | Intensity | 3+ | Nil |
| | 1956 Februart 23 | Solar Flare | Intensity | 3 | Nil |
| | 1956 February 10 | Solar Flare | Intensity | 3 | Nil |
| | 1956 February 17 | Solar Flare | Intensity | 3 | Nil |
| | 1957 September 05 | GM Storm | Dst index in nT | -324 | Nil |
| | 1957 September 13 | GM Storm | Dst index in nT | -426 | Nil |
| | 1957 September23 | GM Storm | Dst index in nT | -302 | Nil |
| | 1958 February 11 | GM Storm | Dst index in nT | -428 | Nil |
| | 1958 July 08 | GM Storm | Dst index in nT | -334 | Nil |
| | 1958 September 04 | GM Storm | Dst index in nT | -305 | Nil |
| | 1959 July 10 | Solar Flare | Intensity | 3+ | Nil |
| | 1959 July 11 | >30 MeV SPE | Fluence (10^9 pr cm-2) | 2.3 | Nil |
| | 1959 July 14 | Solar Flare | Intensity | 3+ | Nil |
| | 1959 July 15 | GM Storm | Dst index in nT | -434 | Nil |
| | 1959 July 16 | Solar Flare | Intensity | 3+ | Nil |
| | 1960 April 30 | GM Storm | Dst index in nT | -325 | Nil |
| | 1960 April01 | GM Storm | Dst index in nT | -325 | Nil |
| | 1960 March 31 | GM Storm | Aa index in nT | 285 | Nil |
| | 1960 Nov 12 | >30 MeV SPE | Fluence (10^9 pr cm-2) | 9 | Nil |
| | 1960 November 05 | Solar Flare | Intensity | 2+ | Nil |
| | 1960 November 10 | Solar Flare | Intensity | 3 | Nil |
| | 1960 November 12 | Solar Flare | Intensity | 3+ | Nil |
| | 1960 November 13 | GM Storm | Dst index in nT | -333 | Nil |
| | 1960 November 15 | Solar Flare | Intensity | 3 | Nil |
| | 1967 May 25 | GM Storm | Aa index in nT | 274 | Nil |
| | 1967 May 26 | GM Storm | Dst index in nT | -391 | Nil |
| | 1972 August 02 | Solar Flare | Intensity | 2B | Nil |
| | 1972 August 04 | GM Storm | Aa index in nT | 280 | Nil |
| | 1972 August 04 | >30 MeV SPE | Fluence (10^9 pr cm-2) | 5 | Nil |
| | 1972 August 07 | Solar Flare | Intensity | 3B | Nil |
| | 1972 August 04 | Solar Flare | Intensity | 3B | Nil |
| | 1978 July 08 | Solar Flare | Intensity | M3,X3 | Nil |
| | 1978 July 10 | Solar Flare | Intensity | M5, X3, M7 | Nil |
| | 1978 July 11 | Solar Flare | Intensity | X15 | Nil |
| | 1980 November 06 | Solar Flare | Intensity | X9 | Nil |
| | 1982 December 17 | Solar Flare | Intensity | X10.1 | Nil |
| | 1982 December15 | Solar Flare | Intensity | X12.9 | Nil |
| | 1982 July 14 | GM Storm | Dst index in nT | -322 | Nil |
| | 1982 July 9 | Solar Flare | Intensity | X9.8 | Nil |
| | 1982 June 06 | Solar Flare | Intensity | X12 | Nil |
| | 1984 April 24 | Solar Flare | Intensity | X13 | Nil |
| | 1984 May 20 | Solar Flare | Intensity | X10.1 | Nil |
| | 1986 February 08 | GM Storm | Aa index in nT | 259 | Nil |
| | 1989 August 16 | Solar Flare | Intensity | X20 | Nil |
| | 1989 March 06 | Solar Flare | Intensity | X15 | Nil |
| | 1989 March 14 | GM Storm | Dst index in nT | -548 | Nil |
| | 1989 October 19 | Solar Flare | Intensity | X13 | Nil |
| | 1989 October 19 | >30 MeV SPE | Fluence (10^9 pr cm-2) | 4.3 | Nil |
| | 1989 September 29 | Solar Flare | Intensity | X9.8 | Nil |
| | 1990 May 24 | Solar Flare | Intensity | X9.3 | Nil |
| | 1991 January 25 | Solar Flare | Intensity | X10 | Nil |
| | 1991 June 04 | Solar Flare | Intensity | X12 | Nil |
| | 1991 June 06 | Solar Flare | Intensity | X12 | Nil |
| | 1991 June 11 | Solar Flare | Intensity | X12 | Nil |
| | 1991 June 15 | Solar Flare | Intensity | X12 | Nil |
| | 1991 June 9 | Solar Flare | Intensity | X10 | Nil |
| | 1991 June01 | Solar Flare | Intensity | X12 | Nil |
| | 1991 March 22 | Solar Flare | Intensity | X9.4 | Nil |
| | 1991 March 25 | GM Storm | Dst index in nT | -297 | Nil |
| | 1991 Novemver 09 | GM Storm | Dst index in nT | -375 | Nil |
| | 1992 November 02 | Solar Flare | Intensity | X9 | Nil |
| | 1997 November 6 | Solar Flare | Intensity | X9.4 | Nil |
| | 2000 July 14 | >30 MeV SPE | Fluence (10^9 pr cm-2) | 4.3 | Nil |
| | 2000 July 16 | GM Storm | Dst index in nT | -301 | Nil |
| | 2000 November 09 | >30 MeV SPE | Fluence (10^9 pr cm-2) | 3.1 | Nil |

**Table 5 List of lunar missions associated with multiple planetary conjunctions during 1957-2000 AD**

| Spacecraft | Launch date | Mission | Mission duration | Country | Status | Conjunction period |
|---|---|---|---|---|---|---|
| E-1,No.3 | 4 Dec 1958 | Impactor | Failed to orbit | Soviet Union | Failure | 1958 Nov 23 – Dec 17 |
| Pioneer 3 | 6 Dec 1958 | Flyby | Failed to orbit | USA | Failure | 1958 Nov 23 – Dec 17 |
| Ranger 8 | 17 Feb 1965 | Impactor | 65 hours | USA | Successful | 1965 Feb 22-Mar 7 |
| AS 201 | 26 Feb 1965 | Space orbitor test flight | | USA | Successful | 1965 Feb 22-Mar 7 |
| Luna 9 | 31 Jan 1966 | Lunar lander | 6 days | Soviet Union | Successful | 1966 Feb 11-Mar 14 |
| KOSMOS 111 | 1 Mar 1966 | Orbitor | Failed to orbit | Soviet Union | Failure | 1966 Feb 11 –Mar 14 |
| Zond 8 | 20 Oct 1970 | Flyby | | Soviet Union | Successful | 1970 Oct 25 – Nov 9 |
| Explorer 49 | 10 Jun 1973 | Orbitor | 2 years | USA | Successful | 1973 May 21 –Jun 12 |
| Luna 24 | 9 Aug 1979 | Lunar sample return | 13 days | Soviet Union | Successful | 1979 Aug 2 - 14 |
| Clementine | 25 Jan 1994 | Orbitor | 115 days | USA | Successful | 1994 Feb 7 - 27 |

**Table 6 List of human space flight missions associated with multiple planetary conjunctions for the period 1961 – 2000 AD.**

| Space craft | Launch Date | Return Date | Date of Conju-nction | Crew |
|---|---|---|---|---|
| Mercury-Atlas(6) | 1962 Feb 20 | 1962 Feb 20 | 1962 Feb 03-Feb 17 | J. H.Glenn |
| Gemini 4 | 1965 Jun 03 | 1965 Jun 07 | 1965 Jun 03-Jun 07 | J. A. Mcdivitt<br>E. H. White |
| Gemini 8 | 1966 Mar 16 | 1966 Mar 17 | 1966 Feb 11-Mar 14 | N. A. Armstrong<br>D. R. Scott |
| Skylab 2 | 1973<br>May 25 | 1973<br>Jun 22 | 1973<br>May 21-<br>Jun12 | C. P. Conrad<br>P. J. Weitz<br>J. P. Kerain |
| Soyuz 38 | 1980<br>Sep 18 | 1980<br>Sep 26 | 1980<br>Aug 30-<br>Sep 17 | Y. Romanenko<br>A. T. Mendez |
| STS-5<br>*Columbia* | 1982<br>Nov 11 | 1982<br>Nov 16 | 1982<br>Nov01-<br>Nov 29 | V. D. Brand<br>R. F. Overmyer<br>J. P. Allen<br>W. B. Lenoir |
| STS-61-B<br>*Atlantis* | 1985<br>Nov 26 | 1985<br>Dec 03 | 1985<br>Nov22-<br>Dec 09 | B. H. Shaw<br>B. D O'Connor<br>M. L. Cleave<br>S. C. Spring<br>J. L. Ross<br>R.Neri Vela<br>C. D. Walker |
| STS-51-L<br>*Challenger* | 1986<br>Jan 28 | | 1986<br>Jan26-<br>Feb23 | F. R. Scobee<br>M. J. Smith<br>J. A. Resnik<br>E. S. Onizuka<br>R. E. McNair<br>G.B. Jasvis<br>S. C McAuliffe |
| STS-30<br>*Atlantis* | 1989<br>May 04 | 1989<br>May 08 | 1989<br>May 06 –<br>May 08 | D. A. Walker<br>R. J. Grabe<br>N. E. Thagard<br>M. L. Cleave<br>M. C. Lee |
| STS-60<br>*Discovery* | 1994<br>Feb 03 | 1994<br>Feb 11 | 1994<br>Feb 07-<br>Feb 27 | C. F. Bolden<br>K. S. Reightler<br>N. Jan Davis<br>R. M. Sege<br>F.R.C. Diaz<br>S.Krikalev |
| STS-101<br>*Atlantis* | 2000<br>May 19 | 2000<br>May 29 | 2000<br>May01-<br>May31 | J D. Halsell<br>S. J. Horowitz<br>M. E. Weber<br>J. N. Williams<br>J S. Voss<br>S. J. Helms<br>Y.Usachev |

**Table 7: Triple conjunctions during 2001-2020 and associated solar-terrestrial conditions**

| Date of closest Triple conjunction | Planets in Conjunction | Solar/Sunspot activity decrease if any | Space weather events associated if any | Space missions associated if any (Human) |
|---|---|---|---|---|
| 2002 May 7 | Ven,Sat,Mar | SSA decrease May5 -15 | GMS: May 11 (-110 nT) | |
| 2005 June 26 | Mer,Sat,Ven | SSA decrease June 17-26 | | |
| 2006 Aug 21 | Mer,Sat,Ven | Nil | | |
| 2006 Dec 10 | Mer,Jup,Mar | SSA decrease Dec 6-18 | GMS (-147nT) Dec 15<br>10 MeV SPE: Dec 13 | ISS Dec 10:Succ |
| 2007 Feb 8 | Mer,Ura,Ven | SSA decrease Feb 3-16 | | |
| 2008 Aug 15 | Mer,Sat,Ven | Nil | Aug 14, 10 MeV: SPE | |
| 2008 Sep 8,12 | Mer,Mar,Ven | Nil | | Sep 25-28: Succ |
| 2009 Feb 24 | Mer,Jup,Mar | Nil | | |
| 2009 April 21 | Ven,Uran,Mar | Nil | | |
| 2009 Oct 9 | Mer,Sat,Ven | Nil | | 11 Oct :Succ |
| 2010 Aug 8 | Ven,Sat,Mar | Nil | | |
| 2011 April 19 | Mer,Jup,Mar | SSA decrease Apr 15-19 | | ISS Apr 4: Succ |
| 2011 May | Mer,Ju,Ve,Mar | Decrease in 10.7cm flux | | |
| 2015 Mar 4 | Ven,Ura,Mar | Nil | | May 16: Succ |
| 2015 Aug 6 | Mer.Jup.Ven | SSA decrease Aug 7-12 | | |
| 2015 Oct 26 | Ven.Jup.Mar | 10.7cm flux decrease<br>Oct 21-26 | Oct 29: 10 MeV SPE | |
| 2016 Aug 28 | | Nil | | |